\documentclass[12pt,a4paper]{article}
\usepackage{amsmath}
\usepackage{epsfig}
\usepackage{subfigure}
\usepackage{amssymb}
\usepackage{colordvi}
\usepackage{cite}

\newcommand{\mec}{|\epsilon|^2}
\newcommand{\Dt}{\Delta t}

\newcommand{\sq}{\frac{1}{\sqrt{2}}}
\newcommand{\w}{\omega}
\newcommand{\Dm}{\Delta m}
\newcommand{\DG}{\Delta \Gamma}
\newcommand{\G}{\Gamma}
\newcommand{\BB}{{|B^0B^0\rangle}}
\newcommand{\BbB}{{|\bar B^0B^0\rangle}}
\newcommand{\BBb}{{|B^0 {\bar B}^0\rangle}}
\newcommand{\BbBb}{{|\bar B^0\bar B^0\rangle}}
\newcommand{\abs}[1]{\left| #1 \right|}
\newcommand{\re}{\frac{Re(\epsilon)}{1+|\epsilon|^2}}
\newcommand{\nn}{\nonumber}
\newcommand{\fig}[1]{Fig.~(\ref{#1})}
\newcommand{\DA}{\Delta A_{sl}}
\renewcommand{\b}{B^0}
\newcommand{\bb}{\bar {B^0}}
\newcommand{\sm}{Standard Model }

\newcommand{\ww}{Weisskopf-Wigner }

\newcommand{\parity}{{\cal P}}
\newcommand{\be}{\begin{equation}}
\newcommand{\bea}{\begin{eqnarray}}
\newcommand{\ee}{\end{equation}}
\newcommand{\eea}{\end{eqnarray}}
\newcommand{\eq}[1]{Eq.\,(\ref{#1})}

\begin{document}

\begin{flushright}
hep-ph/0605211\\IFIC/06-15
\end{flushright}

\begin{center}
\begin{Large}
{\bf $\boldsymbol{\Dt}$-dependent equal-sign dilepton asymmetry and CPTV effects in the symmetry of the $\boldsymbol{\b}$--$\boldsymbol{\bb}$ entangled state}\\
\end{Large}
\vspace{0.5cm}
Ezequiel \'Alvarez$^{(a)}$, Jos\'e Bernab\'eu$^{(a)}$, Miguel Nebot$^{(b)}$\\ \vspace{0.3cm}
{\small \emph{
($a$) Departament de F\'{\i}sica Te\`orica and IFIC,\\ Universitat de Val\`encia-CSIC,\\ E-46100, Burjassot, Spain\\
($b$) Centro de F\'{\i}sica Te\'orica de Part\'{\i}culas (CFTP),\\ Instituto Superior T\'ecnico,\\ P-1049-001, Lisboa, Portugal
}}
\end{center}

\begin{abstract}{In this paper we discuss experimental consequences of a novel kind of CPT violation which would manifest itself in the symmetry of the entangled initial state of $B ^0$ and $\bar B ^0$ through their loss of indistinguishability. The ``wrong'' symmetry component is proportional to $\omega$. We focus our theoretical study on the $\Delta t$-dependence of observables concerning equal-sign dilepton events for which the intensity vanishes at $\Delta t = 0$ in absence of $\omega$.  We find that the charge asymmetry, $A_{sl}$, acquires a $\Delta t$-dependence linear in $\omega$, whose relative importance appears on specific regions of time still unexplored. We also do a statistical analysis for the measurements here proposed and find a high sensitivity to this new CPT-violating effect in this charge asymmetry. We obtain the first limits on the $\omega$-effect by re-analyzing existing data on the $A_{sl}$ asymmetry.}
\end{abstract}

\newpage
\section{Introduction}

Determining the flavour of a single neutral meson, that is \emph{flavour tagging}, is a key ingredient in the interpretation of a variety of experiments such as the ones conducted in $\phi$ or B factories concerning, for example, asymmetries in charge conjugated channels or time dependent mixing of neutral particles. In those cases neutral mesons are or were thought to be produced in perfectly correlated Einstein-Podolsky-Rosen particle-antiparticle states. Under these conditions, if one of the two mesons decays at a given time through a flavour specific channel, the flavour of the second accompanying meson state is automatically infered: this fact has a capital importance in order to develop measures of properties like, for example, C (charge conjugation), P (parity) or CP violation. Underlying these properties the existence of a CPT operator and its consequences, encoded in the (powerful) CPT theorem \cite{CPT}, is central. Despite the extraordinary precision of some experimental tests like the equality of masses of particles and antiparticles \cite{pdg}, CPT violation deserves some attention as it may produce subtle effects. In some theoretical frameworks as, for example, several models of quantum gravity, the basic assumptions on top of which the CPT theorem is constructed may not be fulfilled and hence a proper CPT operator may be ill defined \cite{QG}. It is then appropriate to address the question and study what kind of observable impact could be expected and how it may be revealed in neutral meson experiments. Some efforts along this line appeared in references \cite{prl,plb,nik,KCPTreview,seq,seqb}. 

In this paper we will address additional issues that arise in this context. It is organised as follows: we will first sketch the starting point of this framework, specifically by stating the changes appearing in the initial two neutral mesons state, as produced in B factories, induced by the considered CPT breaking. Then we will study the effects of such a change when the states evolve in time; this will set the conditions required to focus on specific decay configurations like the equal-sign dilepton decays. We will analyse and present a detailed account of what we can expect in observables such as the time dependent asymmetry in the mentioned equal-sign dilepton decays. We will finally use them to obtain some numerical results and to set up bounds for the $\omega$-parameter reflecting the considered fundamental breakdown of CPT invariance.

\section{CPT violation in the $\b\bb$ EPR-correlated state}

In this Section we analyze how the implementation of CPT violation through the breakdown of the particle-antiparticle indistinguishability can modify the initial state of two neutral mesons in the B factories.
In the usual formulations of \emph{entangled} meson states
\cite{cptv,lip68,usual} as produced in the B factories, one imposes the requirement
of {\it Bose statistics} for the state $\b \bb$, which implies that
the physical system must be \emph{symmetric} under the combined operation
of charge conjugation ($C$) and permutation of the spatial
coordinates ($\parity$), $C\parity$.
Specifically, assuming conservation of angular momentum and a
proper existence of the \emph{antiparticle state}, one observes that
for $\b\bb$ states which are $C$ conjugates with $C=(-1)^\ell$ (with
$\ell$ the angular momentum quantum number), the system has to be an
eigenstate of $\parity$ with eigenvalue $(-1)^\ell$.  Hence, for
$\ell=1$  we have that $C=-$,
implying $\parity=-$. As a result the initial correlated state
$\b\bb$ produced in a B factory can be written as
 \be |\psi(0)\rangle_{\w=0} = \sq \left( | \b (+\vec k),\bb(-\vec k)\rangle - |\bb(+\vec k), \b(-\vec k)\rangle \right) ,
 \label{initial}
 \ee
where the  vector $\vec k$ is along the direction of the momenta of the
mesons in the center of mass system.

Among the assumptions underlying the previous paragraph, CPT invariance plays an important role; as above mentioned, and first pointed out in \cite{prl}, a breakdown of CPT invariance is to be expected in some theoretical contexts. With a CPT operator ill defined, particle and antiparticle spaces are somewhat disconnected and hence the requirement of $C\parity=+$ imposed by Bose statistics must be relaxed. As a result, we may rewrite the initial state, 
\eq{initial}, as
 \bea
 |\psi(0)\rangle = \frac{1}{\sqrt{2(1+|\w|^2)}} \times \left\{ | \b (+\vec k),\bb(-\vec k)\rangle - |\bb(+\vec k), \b(-\vec k)\rangle + \right. \nonumber \\
 \left.\w \left( | \b (+\vec k),\bb(-\vec k)\rangle + |\bb(+\vec k), \b(-\vec k)\rangle \right) \right\},
 \label{i}
 \eea
where $\w=|\w| e^{i\Omega}$ is a complex CPT violating parameter, associated
with the non-indistinguishable particle nature of the neutral meson
and antimeson states, which parameterizes the loss of Bose
symmetry.  Observe that in \eq{i} a change in the
sign of $\w$ is equivalent --up to an unimportant global phase-- to the exchange of the particles $\b
\leftrightarrow \bb$.  Moreover, once defined through \eq{i}, the
modulus and phase of $\w$ have physical meaning which could be in
principle measured.

We interpret \eq{i} as a description of a state of two {\it distinguishable} particles at first-order in perturbation theory, written in terms of correlated zeroth-order states with definite permutation symmetry. As is easily seen, the states related by a permutation are different due to the presence of $\w$. Therefore, in order to give physical meaning to $\w$, we \emph{define}, in \eq{i}, $+\vec k$ as the direction of the \emph{first} decay and viceversa, $-\vec k$ as the direction of the second decay, that is, we specify the one-particle states used to construct the collective state. Of course, when both decays are simultaneous, there is no way to distinguish the particular chosen one-particle states, notwithstanding the effects of $\w$ are still present \cite{plb} because Lipkin's argument \cite{lip68} on the vanishing of the amplitude for identical decay channels at $\Dt=0$ no longer operates.

It is clear that the modification of the initial state, \eq{i}, will introduce changes in all the observables at the B factories, since it is the departure point of every analysis.   In order to study the possible modifications of the $\Dt$-dependent observables  we must first evolve in time the new initial state (\eq{i}), then we compute the appropriate intensities for $\Dt$ integrating out the sum of decay times. As we will see, the sole time evolution of the initial state allows to visualize important conceptual changes (see reference \cite{plb}).

\section{Time evolution and conceptual changes due to the
appearance of 'forbidden' states}

To study the time evolution of the initial state in \eq{i} we use the usual \ww formalism. The eigenstates of the effective Hamiltonian operator, the \emph{mass} eigenstates $B_{1,2}$, evolve in time through the effective operator $U(t)$,
$$ U(t)|B_\alpha\rangle = e^{-i\mu_\alpha t} |B_\alpha\rangle \qquad ~\quad (\alpha=1,2). $$

The complex eigenvalues $\mu_\alpha=m_\alpha - i \G_\alpha /2$ include the information on the flavour oscillations as well as on the integrated decay channels.  The relation between these mass eigenstates and the flavour states is given by
\begin{eqnarray}
|B_1\rangle &=& \frac{1}{\sqrt{2\left( 1+|\epsilon+\delta|^2 \right)}} \left[ (1+\epsilon +\delta ) |\b\rangle + (1 - \epsilon - \delta) |\bb \rangle \right] ,\nn \\
|B_2\rangle &=& \frac{1}{\sqrt{2\left( 1+|\epsilon-\delta|^2 \right)}} \left[ (1+\epsilon -\delta ) |\b\rangle - (1 - \epsilon + \delta) |\bb \rangle \right] ,
\end{eqnarray}
where $\epsilon$ and $\delta$ are rephasing variant parameters.  As it is well known, $Re(\epsilon)\neq 0$ implies CP and T violation in the time evolution governed by the effective Hamiltonian responsible of $|\Delta B|=2$ flavour oscillation, whereas $\delta\neq0$ implies CP and CPT violation in this time evolution.  We remark that the CPT violating parameter $\delta$ measures CPT violation in the $B$-oscillation, and is totally independent of the parameter $\w$, whose origin lies in the particle-antiparticle distinguishable nature.

Using this standard time evolution for the direct product of the one-particle-states in \eq{i} we get, up to a global phase, the time-evolved state in the flavour basis:
 \bea
 |\psi(t)\rangle=\frac{e^{-\G t}}{\sqrt{2\left(1+\abs{\omega}^2\right)}}\left\{
 C_{0\bar 0}(t)\BBb + C_{\bar 0 0}(t)\BbB + \right. \nn \\
 \left. C_{00}(t)\BB +C_{\bar 0\bar 0}(t)\BbBb\right\}. \label{t}
 \eea
Here
 \bea
 \begin{array}{rcl}
C_{0\bar 0} &=& 1 + \w \left[ \cosh (\alpha t) + 2 K_\delta ^2  \cosh^2\left(\frac{\alpha t}{2}\right) \right] ,  \\
~ \\
 C_{\bar{0} 0} &=& - 1 + \w \left[ \cosh (\alpha t) + 2 K_\delta ^2  \cosh^2\left(\frac{\alpha t}{2}\right) \right] , \\
~ \\
 C_{00} &=& \w K_+ \left[ -\sinh (\alpha t) + 2 K_\delta \sinh^2
 \left(\frac{\alpha t}{2} \right) \right] , \\
~ \\
 C_{\bar 0 \bar 0} &=& \w K_- \left[ -\sinh (\alpha t) - 2 K_\delta
 \sinh^2 \left(\frac{\alpha t}{2} \right) \right] ;
\end{array}
\label{cs}
 \eea
\bea
 \alpha &=& i\frac{\Dm}{2} + \frac{\DG}{4}, \nn \\
 K_{\pm} &=& \frac{(1\pm\epsilon)^2 - \delta^2}{1 - \epsilon^2 + \delta^2}, \label{micaela bus 170}\\
 K_\delta &=& \frac{2\delta}{1 - \epsilon^2 + \delta^2}, \label{mica2}
 \eea
 $\DG=\G_1-\G_2,\ \Dm=m_1-m_2$ and $\G=(\G_1+\G_2)/2$.

It is worth to notice that phase redefinitions of the single
$B$-meson states such as $\b \mapsto e^{i\gamma} \b ,\, \bb \mapsto
e^{-i\bar \gamma} \bb$ are easily handled through the
transformations of the $K_i$-expressions
  \bea
  K_{\pm} &\mapsto& K_{\pm}e^{\pm i(\gamma-\bar \gamma)} \label{rephase1}\\
  K_{\delta} &\mapsto& K_\delta \label{rephase2} .
  \eea
They lead to explicitly rephasing invariant $C_{0\bar 0}(t)$ and $C_{\bar 0
0}(t)$ coefficients, whereas $C_{00}(t)\mapsto
e^{i(\gamma-\bar\gamma)}C_{00}(t)$ and $C_{\bar 0\bar 0}(t)\mapsto
e^{i(\bar\gamma-\gamma)}C_{\bar 0\bar 0}(t)$ are individually
rephasing-variant, but their dependence on the phase is such that
the considered physical observables are rephasing invariant, as they
should.

Observe how, in passing from \eq{initial} to \eq{i}, the loss of the definite anti-symmetry in the initial state due to $\w$ gives rise to the appearance of the states $\BB$ and $\BbBb$ in the time evolved $|\psi(t)\rangle$, \eq{t}.  These states are \emph{forbidden} at any time if $\w=0$, and they are responsible for the breakdown of the tagging process as usually interpreted.  For instance, a flavour specific decay on one side at time $t_1$ does not determine uniquely the flavour of the other meson at the same time $t_1$.  In fact, due to the same flavour states in \eq{t}, a first flavour specific $\b$ decay filters at that time the state on the other side to be $\sim {\cal O}(1) |\bb\rangle + {\cal O}(\w)|\b\rangle$.  Therefore, the probability of having the same flavour specific decay at the same time ($\Dt=0$) on both sides goes as
 \be
 I(\ell^\pm, \ell^\pm, \Dt=0) \sim |\w|^2 .
 \label{demise}
 \ee
This result shows the breakdown of the most important correlation associated with the EPR entangled state \eq{i} \cite{plb}. It introduces conceptual changes in the analysis, but its observability goes as $|\w|^2$ since for $\Dt=0$ the amplitude to decay into identical channels vanishes due to Bose statistics \cite{lip68}.
 
\section{ Time-dependent observables linear in $\w$}
\label{time dependent}

In this Section we study time dependent observables within the framework of the $\w$-effect; the general feature that we expect is that the interference terms give rise to observables which are linear in $\w$. We analyze correlated flavour specific equal-sign decays on both sides. As a first step\footnote{We could also study effects of $\w\neq 0$ in observables such as the ``golden plated'' channel for the observation of mixing-induced time-dependent CP violation $B^0,\bar B^0\to J/\Psi K_S$, but we focus on those leptonic channels because of their superior statistics.  For effects of the $\omega$ term in $K$ physics, see Ref. \cite{prl}.} we explore the corrections to the equal-sign dilepton intensities and then we study the modification in the behaviour of their asymmetry, $A_{sl}$ .

\subsection{Corrections to the equal-sign dilepton intensity}
Restricting the analysis to equal-sign semi-leptonic final states, the intensity for a first decay $B\to X\ell^\pm$ and a second decay, $\Dt$ later, $B\to X'\ell^\pm$ is written as
 \be I(X \ell^\pm,X' \ell^\pm , \Dt) = \int_0^\infty \left| \langle X \ell^\pm ,X'\ell^\pm |U(t_1) \otimes U(t_1+\Dt) |\psi(0)\rangle \right|^2 dt_1 ,
 \label{int}
 \ee
where $t_1$ is the integrated-out time of the first decay.  The computation of \eq{int} is straightforward using Eq.~(\ref{i}) and the results of the previous Section.  The explicit calculation yields

 {\small
 \bea
 I(X\ell^\pm, X'\ell^\pm, \Delta t) = \frac{1}{8} e^{-\G \Dt} \, |A_X|^2 |A_{X'}|^2 \, \left|\frac{(1+s_\epsilon\, \epsilon)^2-\delta^2/4}{1-\epsilon^2+\delta^2/4}\right|^2 \qquad ~ \ \qquad\ \quad\nonumber \\
 \qquad ~\qquad
 \begin{array}{l}
 \left\{ \Bigg[ \frac{1}{\G} + a_\w \frac{8\G}{4\G^2 + \Dm^2} Re(\w) + \frac{1}{\Gamma} |\w|^2 \Bigg] \cosh \left(\frac{\DG \Dt}{2}\right) + \right. \\
 \Bigg[ - \frac{1}{\G} + b_\w \frac{8\G}{4\G^2 + \Dm^2} Re(\w) -\frac{\Gamma}{\Gamma^2 + \Dm^2} |\w|^2 \Bigg] \cos (\Dm \Dt) +  \\
  \left. \Bigg[ d_\w \frac{4 \Dm}{4\G^2 + \Dm^2}  Re(\w)  + \frac{\Dm}{\Gamma^2 + \Dm^2} |\w|^2 \Bigg] \sin (\Dm \Dt) \right\},
 \end{array}
 \label{piropo2}
 \eea
 }

\noindent where we have neglected subdominant terms like $\w K_\delta$ and $\w \DG$, in order to be consistent we also drop any $\w\delta$ term. The coefficients $s_\epsilon,\ a_\w,\ b_\w$ and $d_\w$ in \eq{piropo2} for the different decays are found in Table \ref{pavese}.
\begin{center}
\begin{table}[h]
\begin{center}
\begin{tabular*}{\columnwidth}{@{\extracolsep{\fill}}|c|c|c|}
\hline
Decay & ~$\left( \ell^+,\ell^+, \Dt \right)$ &  ~$\left( \ell^-,\ell^- ,\Dt \right)$  \\
\hline
{\tiny approx.~transition} & {\tiny ${\cal O}(1+\w)\langle \b |T|\bb \rangle + {\cal O}(\w) \langle \b |T| \b \rangle$} & {\tiny ${\cal O}(1+\w) \langle \bb |T| \b \rangle + {\cal O}(\w) \langle \bb |T| \bb \rangle$} \\
\hline
\hline
$s_\epsilon$   &  1 & -1 \\
\hline
$a_\w$         &  1 & -1 \\
\hline
$b_\w$         & -1 &  1 \\
\hline
$d_\w$         &  1 & -1 \\
\hline
\end{tabular*}
\end{center}

\caption{Sign coefficients in the different intensities in \eq{piropo2}. The second row indicates the transition amplitude for the corresponding single B-processes written in an 'order of magnitude' estimate.}
\label{pavese}
\end{table}
\end{center}

\begin{figure}[h]
\begin{center}
\noindent\subfigure[\label{fig1a}]{\epsfig{file=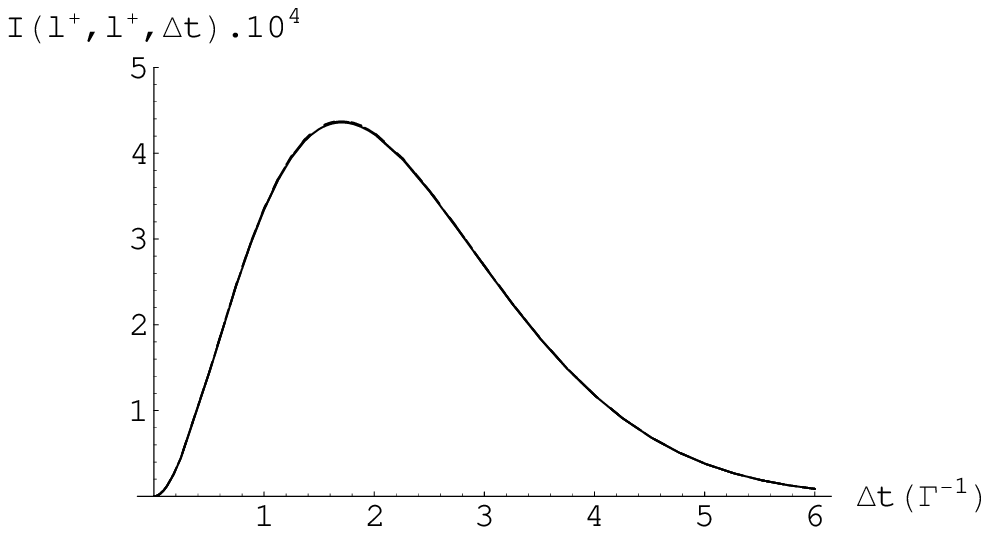,width=0.45\textwidth}}~\subfigure[Zoom.\label{fig1b}]{\epsfig{file=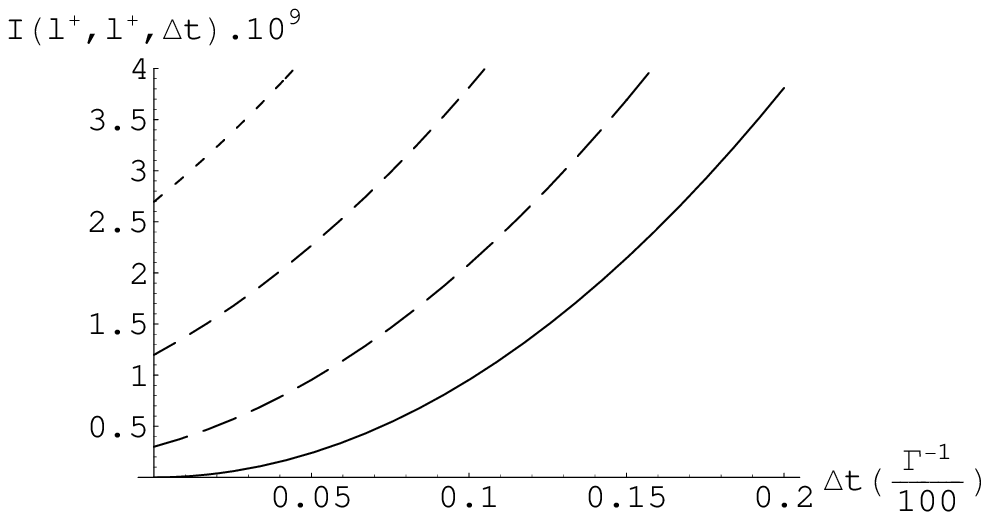,width=0.45\textwidth}}\\
\subfigure[\label{fig1c}]{\epsfig{file=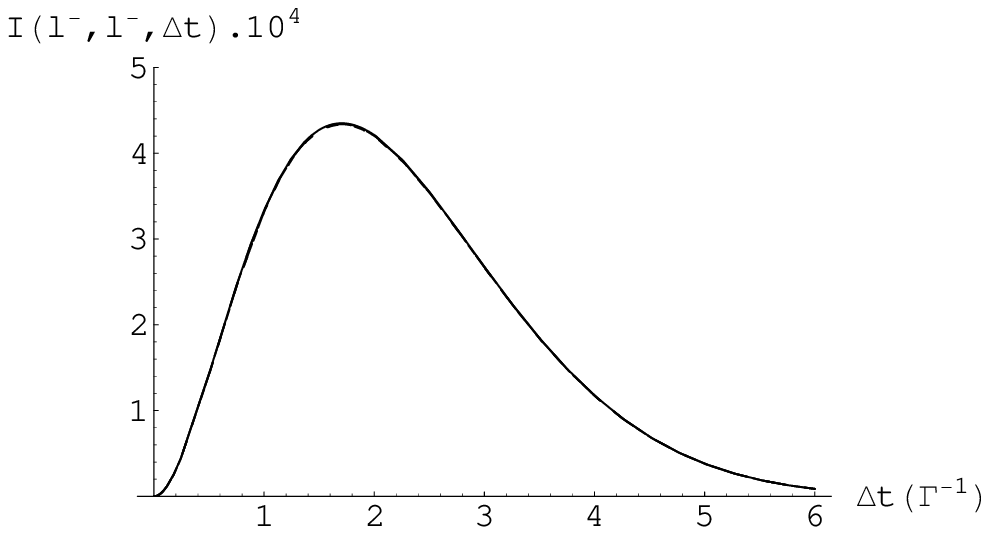,width=0.45\textwidth}}~\subfigure[Zoom.\label{fig1d}]{\epsfig{file=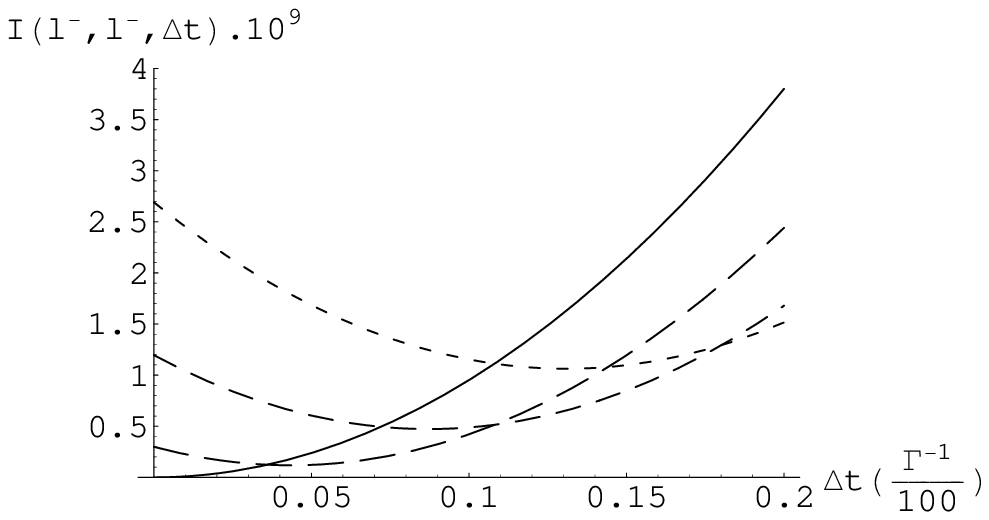,width=0.45\textwidth}}
 \caption{Modification of the equal-sign dilepton intensities due to the presence of $\w$; $\w=0$ (solid-line), $\w=0.0005$ (long-dashed), $\w=0.001$ (medium-dashed) and $\w=0.0015$ (short-dashed).  The zoom figures, $(b)$ and $(d)$, plot the demise of flavour tagging: for $\Dt\to 0$ the intensity goes as $|\w|^2$. Notice also the different behaviour of the intensity as a function of $\w$ in each of the zoom-figures.  As it is shown below, this produces a maximum in their asymmetry for small $\Dt$.}\label{fig1}
\end{center}
\end{figure}

In the following paragraphs we study how the new $\w$-terms in
\eq{piropo2} change the usual $\Dt$-dependent equal-sign dilepton observables.   The dependence of \eq{piropo2} with $\delta$ and $\DG$ is of higher order, so that we can safely present our results in the limit of vanishing $\delta$ and $\DG$.  Effects due to $\w$ are, on the contrary, linear.  In addition we take 
$4Re(\epsilon)/(1+|\epsilon|^2) \approx 1 \times 10^{-3}$ in accordance
with its \sm expected value \cite{pdg}.

The change in the intensity due to the presence of $\w$ is easily
analyzed in \eq{piropo2}, as well as visualized in \fig{fig1}.  As $\G \sim \Dm$, in the region $\Dm\Dt \gtrsim 1$ around the peak, a linear correction in $Re(\w)$ appears in both intensities for the $\ell^+\ell^+$ and $\ell^-\ell^-$ channels.  The relative change of their values is, however, small for small values of $\w$.  On the other hand, observe that in the limit $\Dt \to 0$ the surviving linear terms cancel each other and the leading terms in \eq{piropo2} go as $\sim |\w|^2$.  Although this is a much smaller absolute correction to the intensity, it is worth to study it due to the following reasons:
$(i)$ the modification in the intensity in this case is to be
compared with zero, which is the usual result for $\w=0$; and $(ii)$
it introduces \emph{conceptual changes} in the analysis of the
problem since it constitutes, as mentioned in the previous section,
the breaking of flavour tagging.  Finally, observe the
interesting phenomenon in the intermediate $\Dt$ region, where the
behaviour of the intensity is dominated by the term in the last line
of \eq{piropo2} which is proportional to $\sim d_\w\,
Re(\w)\, (\Dm\Dt)$ and thus linear in $\Dt$.  Therefore we have that, because of $d_w$, an opposite sign behaviour will be seen in each of the intensities (see
Figs.~(\ref{fig1b}) and (\ref{fig1d})): for this sign of $Re(\w)$ we have that
$I(\ell^+,\ell^+,\Dt)$ begins increasing, whereas
$I(\ell^-,\ell^-,\Dt)$ begins decreasing and then grows.  As
analyzed later, this difference between both intensities will produce
a minimum in the denominator of their asymmetry, and hence a
peak for short $\Dt$'s should be expected.  Notice also that this characteristic
behaviour in each of the intensities depends on the sign of $Re(\w)$,
which is measurable.   This was expected, since the initial state is unchanged under the combined exchange $\b\leftrightarrow\bb$ and the sign flip $\w\to -\w$.

\subsection{Behaviour modification in the equal-sign dilepton charge asymmetry, $A_{sl}$}

We now study how the presence of $\w$ in the intensity modifies the
behaviour of the equal-sign dilepton charge asymmetry, also known as
Kabir asymmetry. In the absence of $\w$, flavour-tag by the lepton is operating and this asymmetry becomes a T-violating signal because it compares $\bar B^0\to B^0$ with $B^0\to\bar B^0$.  As it is well known, this asymmetry for
$\w=0$ is of order $Re(\epsilon)$ and \emph{exactly}
time-independent,
 \be
 A_{sl} = \left. \frac{I(\ell^+,\ell^+,\Dt)-I(\ell^-,\ell^-,\Dt)}{I(\ell^+,\ell^+,\Dt)+I(\ell^-,\ell^-,\Dt)}\right|_{\w=0} = 4 \re + {\cal O}(\delta,(Re\ \epsilon)^2),
 \label{ak}
 \ee
  as it can be easily computed by setting $\w=0$ in
\eq{piropo2}. The time independence in \eq{ak} comes from the fact
that when $\w=0$ both intensities have the same time dependence,
which cancels out exactly in the asymmetry.  Notice, however, that $\w\neq 0$ spoils the T-violating property of the asymmetry.  If  $\w \neq 0$ the time dependences of both intensities are not equal any more,
and hence the $A_{sl}$ asymmetry would acquire a time dependence.
Moreover, as discussed below, if $Re(\w) \neq  0$ then the asymmetry
will have an enhanced peak for small $\Dt$, which is quasi-repeated for $\Dm\Dt$ near $2\pi$, and hence there will be optimal regions appropriate to search for experimental evidence of $\w$.  We now analyze separately the different regions in $\Dt$.

In the small $\Dm\Dt$ region one may see qualitatively how the equal-sign dilepton charge
asymmetry produces a peak.  This is easily done by computing the
asymmetry using Eqs.~(\ref{piropo2}) and (\ref{ak}) and keeping in both the numerator and the denominator
terms quadratic in $\Dm\Dt$.  A straightforward analysis of the resulting expression for the regions $\Dm\Dt \ll |\w|$ and $|\w| \ll \Dm\Dt \ll 1$ shows that a peak is expected for the $A_{sl}$ asymmetry in the $\Dm\Dt\sim|\w|$ region.  This peak will be a maximum or a minimum depending on whether $Re(\w)$ is positive or negative, respectively.

\begin{figure}[h]
\begin{center}
\subfigure[$\Omega=0$\label{kabir1}]{\epsfig{file=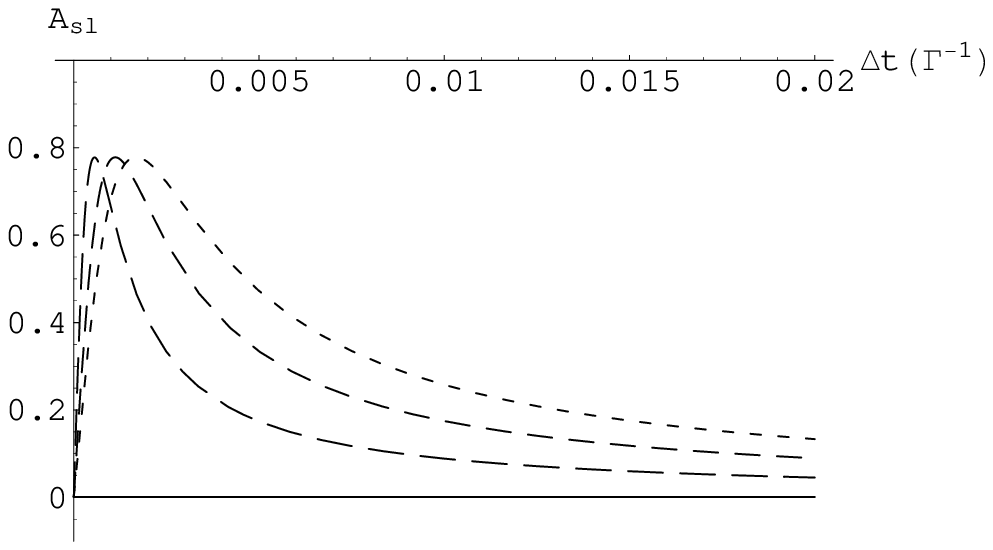,width=0.45\textwidth}}~\subfigure[$\Omega=\frac{3}{2}\pi$\label{kabir2}]{\epsfig{file=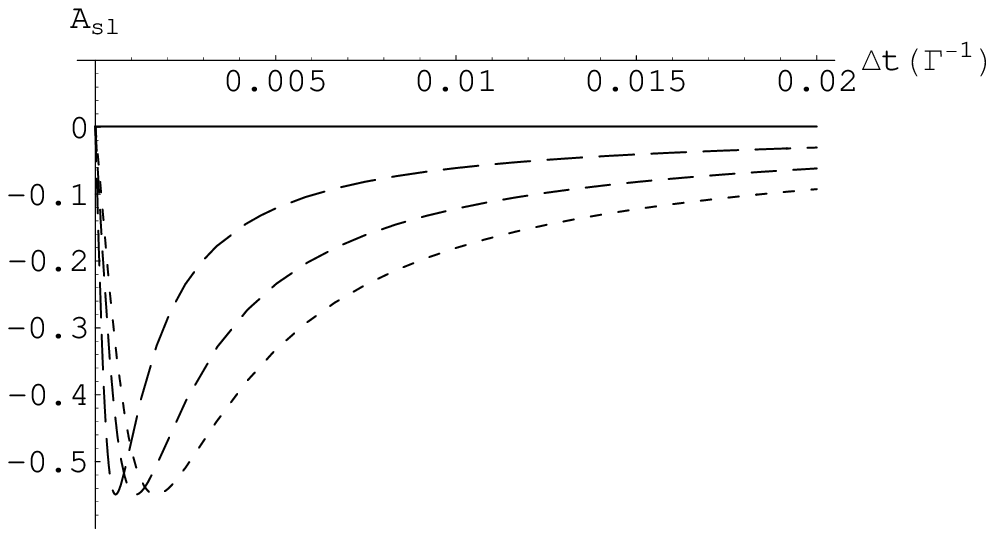,width=0.45\textwidth}}
\caption{Equal-sign dilepton charge asymmetry for different values of $\w$; $|\w|=0$ (solid line), $|\w|=0.0005$ (long-dashed), $|\w|=0.001$ (medium-dashed), $|\w|=0.0015$ (short-dashed).  When $\w\neq0$ a peak of height $A_{sl}(peak) = 0.77 \cos(\Omega)$ appears at $\Dt(peak)=1.12\, |\w| \, \frac{1}{\Gamma}$, producing a drastic difference with the $\w=0$ case, in particular in its time dependence.  Observe that the peak, independently of the value of $|\w|$, can reach enhancements up to $10^3$ times the value of the asymmetry when $\w=0$.}\label{fig2}
\end{center}
\end{figure}


For a better study of the $A_{sl}$ asymmetry in this $\Dm\Dt\ll 1$ region, its plot is shown for different values of $\w$ in \fig{fig2}.  As can be seen, the
asymmetry in the $\w\neq0$ case is not only time dependent but also
has a pronounced peak.  Using \eq{piropo2} one obtains that the peak is
at
 \be
 \Dt_{peak} = \frac{1}{\Gamma} \, \sqrt{\frac{2}{1+x_d^2}} \, |\w| + {\cal O}(\w^2) \approx \frac{1}{\Gamma}\, 1.12 \, |\w| ,
 \label{tmax}
 \ee
where in the last step we have set $x_d\equiv \frac{\Dm}{\G}=0.77$ \cite{pdg}. The
height of the peak is
 \bea
 A_{sl}(\Dt_{peak}) &=& \frac{\cos(\Omega) + \sqrt{2} \, \frac{4+x_d^2}{\sqrt{1+x_d^2}} \, \re}{4\, \cos(\Omega) \re + \frac{1}{2\sqrt{2}} \, \frac{4+x_d^2}{\sqrt{1+x_d^2}}} + {\cal O}(w, \, (Re\ \epsilon)^2)  \nonumber \\
 &=& 0.77 \cos (\Omega) + {\cal O}(\w, \, Re(\epsilon))
 \label{hmax}
 \eea
 We see that although the position of the peak is linearly dependent
on the absolute value of $\w$ (\eq{tmax}), \emph{its height is independent of the
moduli of $\w$ and only depends on its phase} (\eq{hmax}).  Moreover, the
enhancing numerical factor $0.77$ in the leading term of the height
of the peak, \eq{hmax}, is about three orders of magnitude
larger than the expected value of the asymmetry if $\w=0$.  If statistics was enough and the background controllable, this region of $\Dt$ is then clearly the best to measure $\w$ from the asymmetry:  it induces a pronounced time-dependence and its value is controlled by the modulus and phase of $\w$.

In the region $\Dm\Dt \sim 1$ we expect a smooth behaviour with a little time-dependence of $A_{sl}(\Dt)$ for small $\w$.  In fact, analyzing the
time-derivative at the typical time $\Dt=1/\G$ we find
\bea
\frac{1}{\G}\,\frac{d\, A_{sl}}{d\, \Dt} (\Dt=1/\G) = -0.78 Re(\w) +{\cal O}(\w^2),
\label{pavesetrola}
\eea
a result which is independent of $\epsilon$ and $\DG$ (in its
allowed range $\DG/\G\sim10^{-3}$).  However, a shift with respect to
the $\w=0$ case will be always present.  The calculation of the
asymmetry at this time gives an estimate of the expected shift:
\bea
A_{sl}(\Dt=1/\G) &=& \frac{4Re(\epsilon)}{1+\mec} + 3.39 Re(\w) + {\cal O}(\w^2).
\label{shift}
\eea

\begin{figure}[h]
\begin{center}
\subfigure[\label{moco}]{\epsfig{file=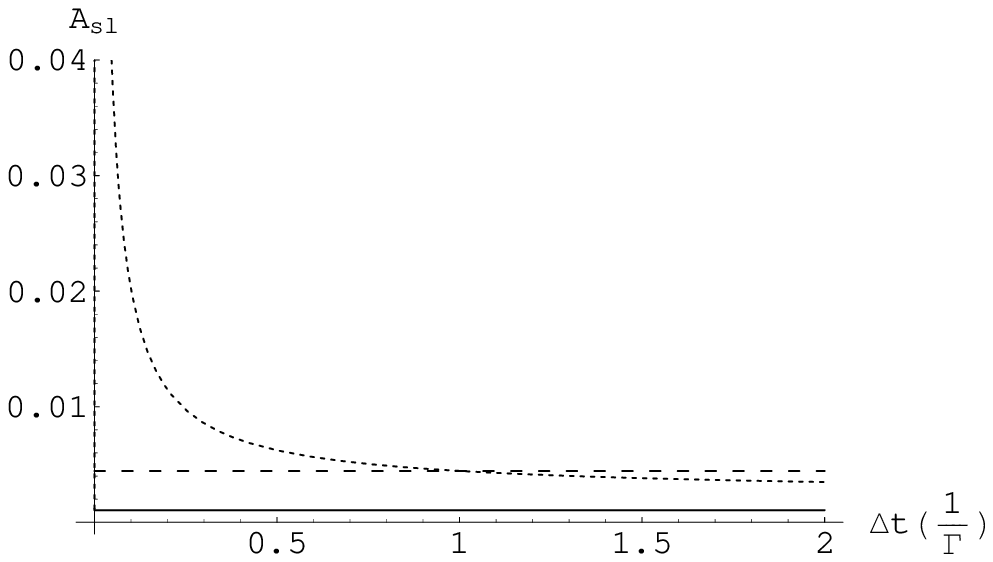,width=0.45\textwidth}}~\subfigure[\label{moco2}]{\epsfig{file=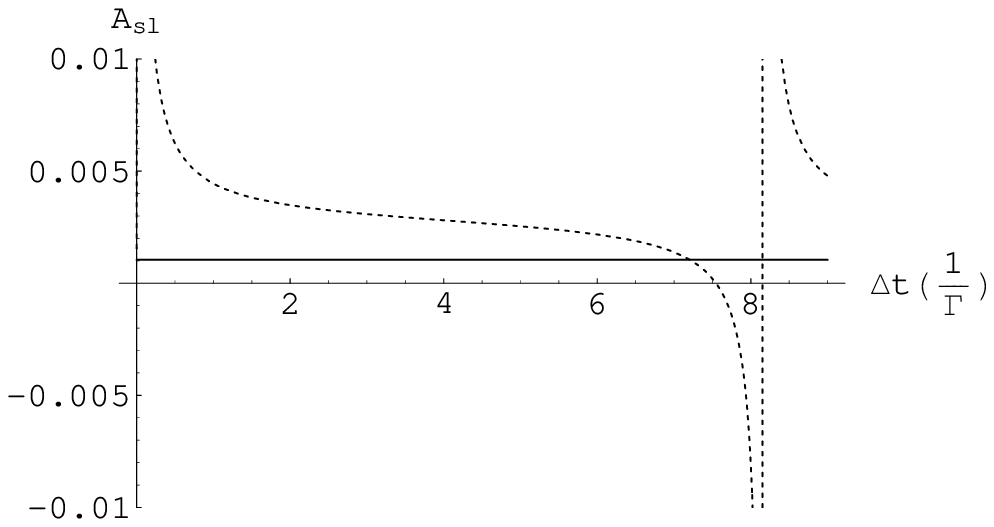,width=0.45\textwidth}}
\caption{The $A_{sl}(\Dt)$ asymmetry at times $\Dt \gtrsim 1/\G$. The solid line represents the $\w=0$ case, the short-dashed line is for $\w=0.001$, and the long-dashed line (in Fig.~(a)) is the shift-approximation for this range of times according to \eq{shift}. In Fig.~(a) we represent the region of times where the asymmetry is quasi time-independent but shifted due to $\w$.  In Fig.~(b) we plot it in a range including $\Dm\Dt \sim 2\pi$ to show the second peak, due to the quasi periodicity of the asymmetry.}
\end{center}
\end{figure}

From this discussion we see that for small enough $\w$, in the range
of times around $\G^{-1} \lesssim  \Dt \lesssim 2\pi/\Dm$, a good approximation for the
shift in $A_{sl}$ goes linear in $Re(\w)$, as \eq{shift} shows.  This shift-effect is displayed in \fig{moco}.

In the region $\Dm\Dt\sim2\pi$ the asymmetry is expected to repeat the small $\Dm\Dt$ region peak-behaviour.  In fact, the asymmetry
is constructed from the intensities in \eq{piropo2}, where the only
non-periodical terms are those that go with $\cosh(\DG\Dt/2)$, but
these terms are quasi-constant for small enough
$\DG$.  In virtue of this, it is expected that the asymmetry repeats its behaviour when the sine and cosine's period is reached. This means that the
above-studied peak-behaviour for small $\Dt$ will be found again at times
around $\Dt \sim2\pi/\Dm = 8.2\G^{-1}$; this time accompanied by
a counter peak with opposite-sign just before $\Dm\Dt=2\pi$ (see \fig{moco2}).  Notice, however, that at these times the available amount of events is suppressed by a factor $e^{-8.2}\sim 10^{-4}$. In any case we point out that, although with still low statistics and low time resolution, the available data is beginning to explore this region \cite{Nakano:2005jb}. (We note that the suppression of this second peak due to $\DG$ is hardly noticeable for $\DG/\G \leq 0.005$.)

\section{Using the $A_{sl}$ asymmetry to observe or constrain the $\w$-effect}
\label{por fin}

From the discussion in the previous Section, we find that there could be good perspectives to measure a non-zero value of $\w$ in the $A_{sl}$ asymmetry.  In this Section we analyze how likely these experimental perspectives are, and at last we put indirect limits on $\w$ using the available data on the dilepton charge asymmetry.

\subsection{Perspectives to observe $\w$ in the $A_{sl}$ asymmetry}

\begin{figure}[h]
\begin{center}
\subfigure[$|\w|\ vs.\ \Dt(\G^{-1})$; for $\Omega=0$\label{level1}]{\epsfig{file=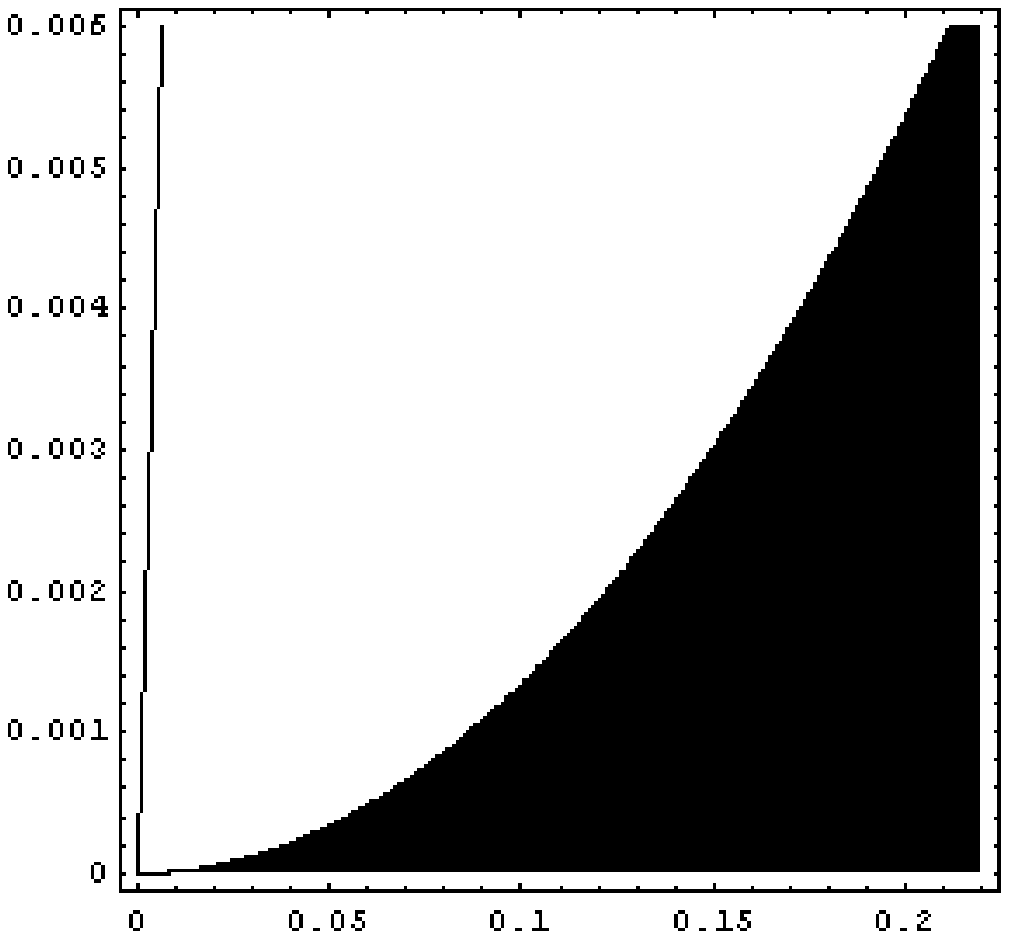,width=0.45\textwidth}}~\subfigure[$\Omega\ vs.\ \Dt(\G^{-1})$; for $|\w|=0.001$ \label{level2}]{\epsfig{file=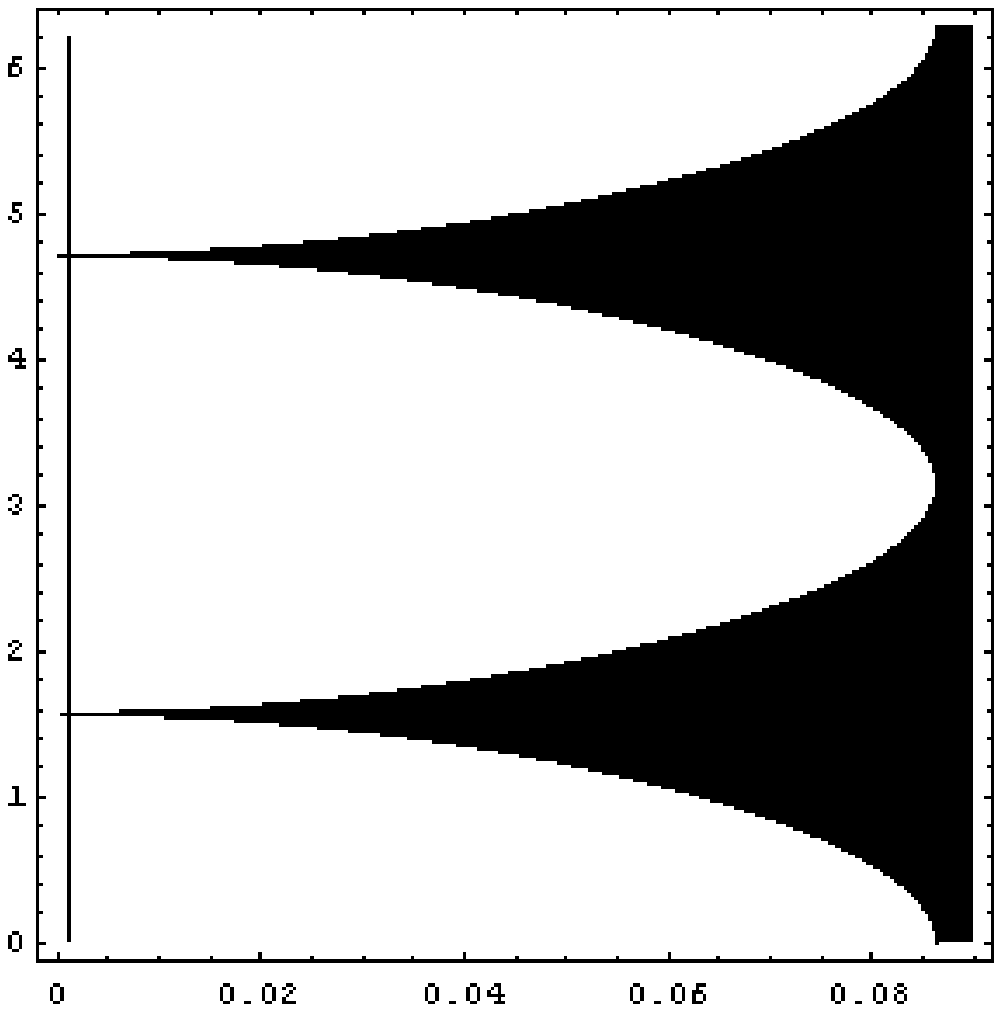,width=0.45\textwidth}}
\caption{Contour curves for $\frac{1}{\G}|dA_{sl} / d\Dt |= 0.1$, the white area represents the points where $\frac{1}{\G}|dA_{sl} / d\Dt | > 0.1$, and hence the time variation would be (expected to be) experimentally detectable. Notice the tiny dark line on the left of each graph which represents the first peak of the asymmetry, where of course the derivative also goes to zero.  Fig.\ (a) plots $|\w|\ vs. \ \Dt$ for a fixed $\Omega=0$, observe that although to see the peak in $A_{sl}$  a very high $\Dt$-resolution is required, the region where the time variation is detectable might be more accessible experimentally. Fig.\ (b) plots the phase $\Omega \ vs.\ \Dt$ for a fixed value of $|\w|=0.001$, note that disregarding the values of the phase around $\pi/2$ and $3\pi/2$, the measurable region (white) is quite favoured in $\Dt$.}\label{fig3}
\end{center}
\end{figure}

In order to detect the $\w$-effect in the $A_{sl}$ asymmetry we rely on the main difference between the $\w=0$ and $\w\neq0$ case: the strong time-dependence.  We are aware, however, that detecting the peaks in $A_{sl}$ may require a fantastic time-resolution.  Therefore, we focus our analysis in the detection of the time-dependence provided by the tail of the peaks.  In this case we may define, with operative purposes, a criterion of
experimental observability regarding the value of the time
derivative $d A_{sl} /d \Dt$.  Considering the detection of the small-$\Dt$ peak --the reasoning for the second peak is analogous--, we may use a limit-detectable-time $\Dt_{limit}$ such that
 \be
 \frac{1}{\G}\left| \frac{d A_{sl} }{d \Dt}(\Dt_{limit} ) \right|= 0.1,
 \label{carlinha}
 \ee 
and hence for $\Dt < \Dt_{limit}$ it would be possible to observe the effect of $\w$ as a time-dependence in the $A_{sl}$ asymmetry. In
\fig{level1} the contour curve is plotted for $\frac{1}{\G}|d A_{sl} /d
\Dt| = 0.1$ as a function of $|\w|$ and $\Dt$ for $\Omega=0$.  As it
can be seen, a value for instance of $\w \sim5\times 10^{-3}$ gives a
$\Dt_{limit} \sim 0.2 \Gamma^{-1}$ as compared to
$\Dt_{peak} \approx 0.005\G^{-1}$.  We conclude that, in much later $\Dt$'s
than $\Dt_{peak}$ the time dependence is still detectable.  In
\fig{level2} the same contour curve is shown, but now plotted as a
function of $\Omega$ and $\Dt$, whereas the modulus is set to $|\w|=
0.001$.  In the figure it can be seen that, disregarding the
values of $\Omega$ close to $\pi/2$ or $3\pi/2$, the measurement of
the equal-sign charge asymmetry $A_{sl}$ at small $\Dt$ represents a
quite good observable to test the $\w$-effect.

\begin{figure}[h]
\begin{center}
\epsfig{file=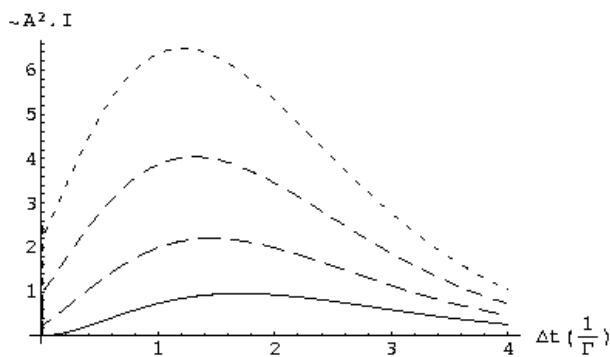,width=0.6\textwidth}
\caption{Figure of merit for measuring the equal-sign dilepton charge asymmetry; from bottom to top $\w=0$ (solid line), $\w=0.0003$ (long-dashed), $\w=0.0006$ (medium-dashed) and $\w=0.001$ (short-dashed).  Observe how the peak in the $A_{sl}$ asymmetry for $\w\neq0$ is reflected here as a growth and a $\Dt$-shift towards the origin of the maximum. This shows that as $\w\neq0$, its effects are found to be in the small $\Dt$ region {\it and at the same time} this region becomes statistically easier to explore. }\label{fig4}
\end{center}
\end{figure}

To complete the analysis of a possible measurement of the $\w$-effect through the equal-sign dilepton charge asymmetry, we study how the statistics requirements to measure $A_{sl}$ are modified due to the presence of $\w$. 
(Notice that although the quantity $dA_{sl}/d\Dt$ is more directly connected to the detection of $\w$, this quantity is not measured, but instead is fitted from the points of $A_{sl}$ for different $\Dt$'s.)
From the statistical analysis point of view, the relevant quantity to observe an asymmetry at different $\Dt$'s is its figure of merit, i.e.\ $I(X,X',\Dt)\cdot (A_{sl}(\Dt))^2$.  As it is well known, the minimum number of events needed to measure a non-compatible-with-zero value of the asymmetry is proportional to its inverse.  The figure of merit is plotted in \fig{fig4} for different values of $\w$.  In this plot one may see the sensitivity that exists to $\w$: as $|Re(\w)|$ grows, the maximum also grows and it is shifted towards smaller $\Dt$'s.  This behaviour gives the asymmetry a high sensitivity to $\w$: as was explained above, the effects of $\w$ are found enhanced in the small $\Dt$ region, whereas the figure of merit tells us that as $|Re(\w)|$ grows, the exploration of this region becomes more effective.

\subsection{Present limits on $\w$ using existing data}

Although the best region to see the effects of $\w$ is at small
$\Dt$ and at $\Dt\approx 2\pi/\Dm$, the existing data on the $A_{sl}$ asymmetry have not been
taken effectively in those regions yet.  The explored region, $0.8\G^{-1} \lesssim \Dt \lesssim 10\G^{-1}$ \cite{Nakano:2005jb, Aubert:2002mn}, has left out the first peak and its tail, whereas the region of the second peak is poor in statistics, since it is suppressed by a factor of $10^{-4}$.   In any case, this existing data could be used to put limits on $\w$, since its existence would introduce a time dependence and produce a shift in the $\Dt\sim 1/\G$ region, as shown in \eq{shift}.

We point out here that putting limits to $\w$ using the available fit
of $A_{sl}$ to a constant is an indirect constraint on $\w$.  It is clear that a direct constrain on $\w$
should be obtained using the original complete experimental data and fitting it to
the expression $A_{sl}(\w,\Dt)$ that comes out from constructing the
asymmetry using \eq{piropo2} with its time dependence.

In order to find indirectly the allowed values for $\w$ using the existing
measurements of $A_{sl}$ we proceed as follows. Using \eq{piropo2}
we compute the asymmetry as a function of $\w$, and we allow
$\epsilon$ to vary within its \sm expected range \cite{pdg},
 \bea
 \left| \frac{4Re(\epsilon)}{1+\mec}\right| = {\cal O}\left( \frac{m_c^2}{m_t^2} \sin(2\beta) \right) \lesssim 0.001 .
 \label{el che}
 \eea
 The values of $\w$ which are within the $95\%$ C.L. region are those such
 that $A_{sl}(\Dt,\w)$ is within the two standard deviation of the measured
value of the asymmetry,
 \bea
 A_{sl}^{exp} = 0.0019 \pm 0.0105 .
 \eea
(This value comes from the equally weighted average of the results of Refs.\
\cite{Nakano:2005jb,Aubert:2002mn}.)  Of course that we
need to define what is {\it 'within'}, since $A_{sl}(\Dt,\w)$ is
time-dependent, whereas $A_{sl}^{exp}$ is constant and, besides, the
density of experimental events, $\rho(\Dt)$, is highly
$\Dt$-dependent.  With this purpose, we define a density-weighted average
difference between the experimental and $\w$-theoretical asymmetries
as
 \bea
 \DA (\w) &=& \frac{\int_{\Dt_1}^{\Dt_2} \rho(\Dt)(A_{sl}(\Dt,\w)-A_{sl}^{exp})\, d\Dt}{\int_{\Dt_1}^{\Dt_2}\rho(\Dt)\, d\Dt} ,
 \label{pendorcha}
 \eea
 where $(\Dt_1,\Dt_2)=(0.8\frac{1}{\G},8\frac{1}{\G})$ is the maximum common range of times where the measurements have been performed, and the density of events is
 \bea
 \rho(\Dt) &=& e^{-\G\Dt} (1-\cos(\Dm\Dt)).
 \label{que conchita me chupe}
 \eea
In the density function $\rho(\Dt)$ we
have approximated $\DG\approx\w\approx0$ since at this point they
constitute second order contributions. From the definition of
the weighted-average-difference in \eq{pendorcha} we obtain the $95\%$
C.L.\ allowed values for $\w$ as those that furnish
 \bea
 | \DA (\w) | \leq 2\sigma = 0.0210 .
 \eea
 The fulfilment of this equation, neglecting quadratic contributions in $\w$, gives the final result:
 \bea
 -0.0084 \leq Re(\w) \leq 0.0100 \qquad ~ \quad 95\%\mbox{C.L.}
\label{pitufina}
 \eea
  Where the values of $\epsilon$ used to define the upper and
 lower limits correspond in each case to those in \eq{el che} which
 extend the most the allowed range for $\w$.  These are the first
 known limits on $\w$ \cite{seqbis}.

It is interesting to notice that, given the range of times where the integral in \eq{pendorcha} is performed,  we could repeat the same calculation but approximating $A_{sl}(\Dt,\w)$ by its value at the typical time $\Dt=1/\G$ (see \eq{shift}).  In this case we obtain the approximated limits $-0.0059\lesssim Re(\w) \lesssim 0.0070$ which are a good --and simpler-- {\it estimate} compared to the exact value in \eq{pitufina}.

\section{Conclusions}
In this article we have studied the possible consequences that CPT violation through the loss of indistinguishability of particle-antiparticle may have in some of the usual observables of the B-factories.  We have shown that the modification of the initial $\b\bb$ entangled-state due to this $\w$-effect could bring out important conceptual and measurables changes in the correlated B-meson system.

In order to find linear-in-$\w$ observables we have studied the $\Dt$-dependent observables.  In particular, we have addressed our attention to the equal-sign dilepton charge asymmetry, $A_{sl}$, which we have found to be an encouraging observable to look for experimental evidence.  As it is well known, the $A_{sl}$ asymmetry in the \sm is expected to be constant.  On the other hand, in the $\w\neq 0$ case we have found that the asymmetry acquires a strong time dependence for short and large $\Dt$'s through two peaks, wheras in the intermediate region -- where it has been measured -- the behaviour is smooth with a little time dependence.  

We have shown that the measurement of $\w$ should be seeked by either experimentally exploring the regions where the time-dependence of $A_{sl}$ is strong, or either by measuring the smooth time dependence in the intermediate region where the figure of merit for the asymmetry is higher.  In the latter case, we have investigated on the possibility of measuring the time-dependence through the tail of the peaks, and we have proved that these tails broaden considerably the measurable region.  As a first manifestation of this analysis, we have set the first limit on $\w$ using the available experimental data.   We have found $|\w| \lesssim 10^{-2}$ at a $95\%$ confidence limit.  A re-analysis of the complete experimental data would lead straightforwardly to more stringents bounds on $\w$.

\section*{Acknowledgments}
The authors are indebted to G. Isidori, N. Mavromatos and J. Papavassiliou for most interesting discussions. The research has been supported by Spanish MEC and European FEDER under grant FPA 2005-01678, and by \emph{Funda\c{c}\~{a}o para a Ci\^{e}ncia e a Tecnologia} (FCT, Portugal) under project CFTP-FCT UNIT 777. E.A. acknowledges financial support from Universitat de Val\`{e}ncia; M.N. acknowledges financial support from FCT.


\end{document}